\documentclass[aps,showpacs,floatfix,twocolumn,prl]{revtex4}

\usepackage{amsmath}
\usepackage{amssymb}
\usepackage{graphicx}
\usepackage{tabularx}

\begin{document}

\title{Emergent geometries and nonlinear-wave dynamics in photon fluids}

\author{F. Marino$^{1,2}$, C. Maitland$^3$, D. Vocke$^3$,  A. Ortolan$^4$, D. Faccio$^3$.}
\email{marino@fi.infn.it, d.faccio@hw.ac.uk}
\address{$^1$ CNR-Istituto Nazionale di Ottica, L.go E. Fermi 6, I-50125 Firenze, Italy \\
$^2$ INFN, Sez. di Firenze, Via Sansone 1, I-50019 Sesto Fiorentino (FI), Italy\\
$^3$ School of Engineering and Physical Sciences, SUPA, Institute of Photonics and Quantum Sciences Heriot-Watt University, Edinburgh EH14 4AS, UK\\
$^4$ INFN, Laboratori Nazionali di Legnaro, Viale dell’Universita 2, I-35020 Legnaro (PD), Italy}

\date{\today}

\begin{abstract}
Nonlinear waves in defocusing media are investigated in the framework of the hydrodynamic description of light as a photon fluid.
The observations are interpreted in terms of an emergent curved spacetime generated by the waves themselves, which fully determines their dynamics. The spacetime geometry emerges naturally as a result of the nonlinear interaction between the waves and the self-induced background flow.
In particular, as observed in real fluids, different points of the wave profile propagate at different velocities leading to the self-steepening of the wave front and to the formation of a shock. This phenomenon can be associated to a curvature singularity of the emergent metric. Our analysis offers an alternative insight into the problem of shock formation and provides a demonstration of an analogue gravity model that goes beyond the kinematic level.
\end{abstract}


\maketitle
Geometry plays a fundamental role in the description of disparate phenomena across different areas of physics, ranging from continuous mechanics and nonlinear dynamics to electromagnetism and high-energy physics. The paradigmatic example is General Relativity (GR) where, far from being merely a convenient representation, the language of differential geometry allows us to capture the essence of gravitational force.
A recent active field of research in which geometrical concepts naturally comes into play is analogue gravity (for a review, see \cite{rev}). 
The general idea is that the propagation of a scalar field on a curved spacetime can be reproduced in condensed-matter systems by studying the evolution of elementary excitations on top of a suitable background configuration. For instance, sound waves in flowing fluids propagate in an effective Lorentzian geometry (acoustic metric) which is determined by the physical properties of the flow \cite{unruh,visser}. Hence, by properly choosing the background flow, it is possible to mimic black holes and several aspects of quantum field theory in curved spacetime. \\
These studies are generally based on perturbative schemes, i.e. they consider the propagation of linearized fluctuations over a given background solution of the full nonlinear problem. As such, the analogy works only at the kinematical level: the geometrical description holds only for the fluctuations, while the background couples with the usual flat metric and is not affected by the perturbation dynamics.\\
 A new perspective was opened by Goulart in Ref.~\cite{novello}. Under certain conditions, the nonlinear dynamics of a scalar field can be described in terms of an emergent spacetime geometry, which is generated by the field itself and determines its propagation. This result  extends the analog gravity approach as the emergent metric includes the whole dynamics of the background field and is not restricted to linearized fluctuations. 
Remarkably, a similar situation occurs in GR: the metric on which a given scalar field propagates is modified by the scalar field itself. \\
We note that the emergent metric in nonlinear models does not depend on the scalar field via Einstein's equations and thus they cannot be directly considered as models for Einstein gravity. Nevertheless, they encode in a geometric framework the dynamical interplay between the scalar field and the metric and so they could provide valuable insights into relevant dynamical features present also in GR such as back-reaction effects at a full nonlinear level \cite{br}.\\
Compressible, non viscous fluids appear as the ideal candidates for this kind of investigation and in principle, can push the analogy further. In the linear regime, they are the prototype model for studies on acoustic black holes \cite{rev}, Hawking radiation \cite{unruh,visser} and superradiance \cite{schutz,basak,berti}. Extending to the nonlinear regime, we find that causality is still governed by an acoustic metric \cite{white,chrusciel} 
and the evolution of nonlinear perturbations can be described through the analog gravity formalism \cite{cherubini}. Acoustic geometries which dynamically depend on the fluid variables appear also in the context of black-hole accretion models \cite{moncrief,ray1} and spherically symmetric outflows of nuclear matter \cite{ray2}. Even more importantly, a precise correspondence exists between the gravitational field equations and suitably constrained nonlinear flows \cite{cadoni2}. Therefore such systems could be exploited to mimic some aspects of black holes dynamics and spacetime singularities \cite{cadoni}.\\
As an alternative to actual flowing fluids, light beams propagating in defocusing optical media are promising models for analogue gravity experiments \cite{vocke,elazar,marino,fouxon,carusotto}.
The optical field dynamics can be described as a fluid of interacting photons \cite{rica,chiao} on which linear perturbations, i.e. sound waves, experience an effective curved spacetime determined by the field intensity and phase pattern. \\
In this Letter, we report an experimental study of the dynamics of \emph{nonlinear} acoustic disturbances in such photon fluids and we provide a geometrical interpretation of the results. We show that nonlinear density waves induce a background flow which, in turn, strongly modifies their propagation. In analogy with GR, this self-interaction is interpreted in terms of an effective curved geometry generated by the wave itself. In particular, as observed in compressible, non viscous fluids, different points of the wave profile propagate at different velocities leading to the self-steepening of the wave front and the subsequent formation of a shock. This phenomenon is associated to a curvature singularity of the emergent metric.\\
{\emph{Photon-fluid model:}} We consider the propagation of a monochromatic optical beam of wavelength $\lambda$, in a 1D defocusing medium. The slowly varying envelope of the optical field follows the Nonlinear Schr{\"o}dinger Equation (NSE)
\begin{equation}
\partial_z E = \frac{i}{2k}\partial_{xx}^2 E - i \frac{k}{n_0} E \Delta n\;
\label{eq1}
\end{equation}
where $z$ is the propagation direction, $k = 2 \pi n_0 / \lambda $ is the wave number and $\partial_{xx}^{2} E$, defined with respect to the transverse coordinate $x$, accounts for diffraction. The term $\Delta n(\vert E \vert^{2}, x, z)$ describes the self-defocusing effect and in the case of thermo-optical nonlinearities can be highly nonlocal, i.e. the change in refractive index at any position depends not only on the local intensity, but also on surrounding field intensities \cite{vocke}.\\
When nonlocal effects are negligible, $\Delta n = n_2 \vert E \vert^{2}$ and, in the Madelung formulation $E = \rho^{1/2} e^{i \phi}$, Eq. (\ref{eq1}) can be recast into the continuity and Euler equation for a 1D-Bose-Einstein condensate 
\begin{eqnarray}
\partial_t \rho + \partial_{x}(\rho v) = 0 \label{eq2} \; \\
\partial_t \psi + \frac{1}{2} v^2 + \frac{c^2}{n_0^3}n_2 \rho - \frac{c^2}{2 k^2 n_0^2}\frac{\partial_{xx}^2 \rho^{1/2}}{\rho^{1/2}} = 0 \label{eq3}\;  
\end{eqnarray}
where the optical intensity $\rho$ represents the fluid density, $v = \frac{c}{k n_0}\partial_{x} \phi$ is the fluid 
velocity ($c$ is the speed of light) and the propagation direction has been considered as a time coordinate $t = \frac{n_0}{c} z$. In Eq. (\ref{eq3}), the optical nonlinearity $\frac{c^2}{n_0^3}n_2 \rho$ corresponds to the repulsive atomic interactions, while the last term is the analog of a quantum pressure. 
By applying the $\partial_{x}$ operator to Eq. (\ref{eq3}) and neglecting the quantum pressure term, we obtain 
\begin{equation}
\partial_t v + v \partial_x v + \frac{\partial_{x} P}{\rho}= 0 \, \\
\label{eq4}
\end{equation}
where we have defined the bulk pressure $P =\frac{c^2 n_2}{2 n_0^3} \rho^{2}$. \\
Equations (\ref{eq2})-(\ref{eq4}) are exactly the Navier-Stokes equations in a 1D compressible, non viscous fluid. The nonlinear term, $v \partial_x v$, determines the steepening rate of a wave and will lead to the formation of a shock front \cite{landau,courant,lax} that in real fluids will be smoothed out by diffusion processes or viscosity and, in photon fluids, by the quantum pressure and nonlocal terms. 
\begin{figure}[!t]
\begin{center}
\includegraphics[width=8cm]{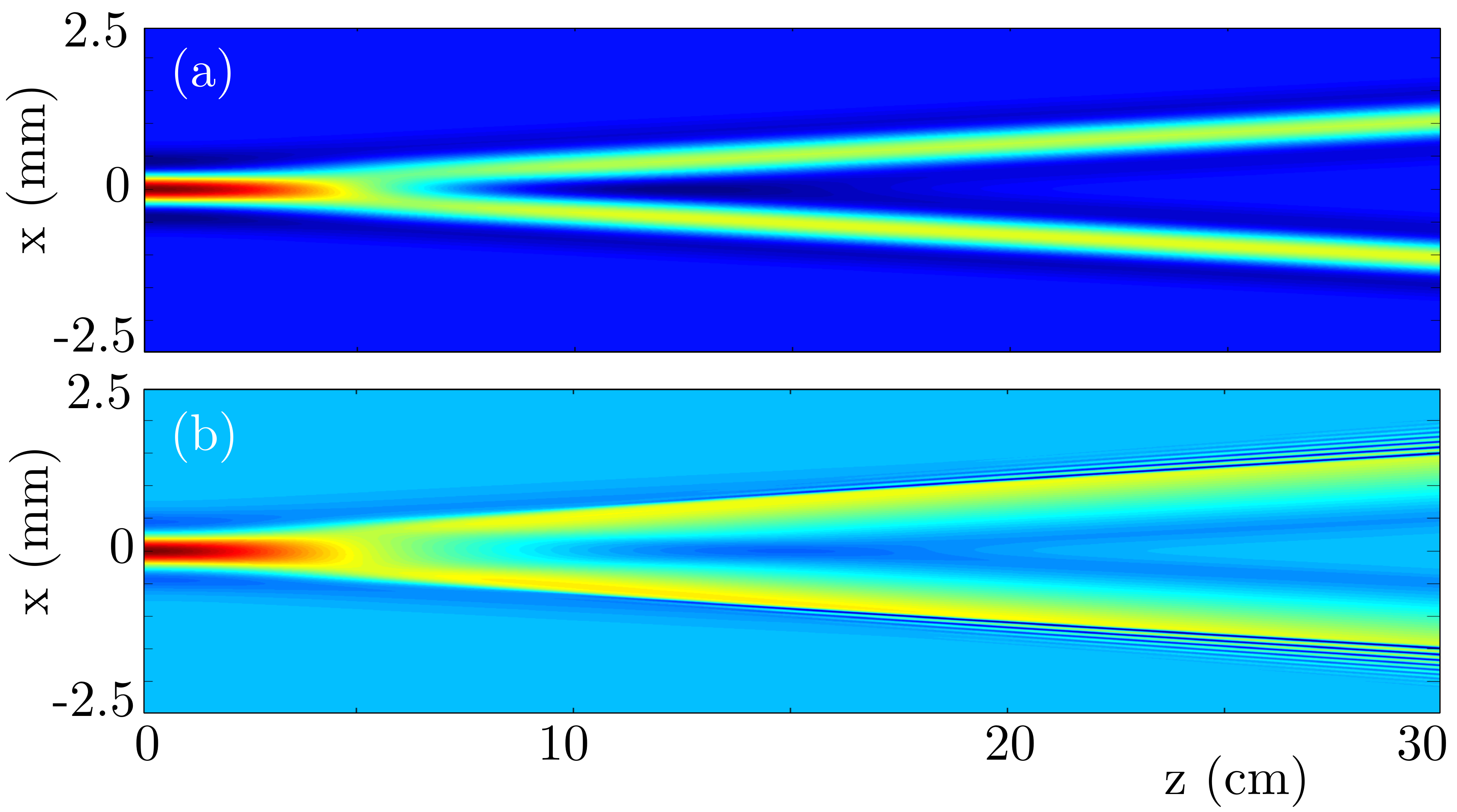}
\caption{Numerical results: probe beam evolution in the presence of a flat-top pump beam. (a) A weak probe beam (1\% of the pump power, 0.3 deg input angle) splits into two beams corresponding to two counterpropagating Bogoloiubov particles or density waves in the fluid. (b) At high probe intensities (same power as the pump) the density waves propagate nonlinearly and develop (supersonic) shock fronts that form ripples on the leading edges.  \label{fig1} }
\end{center}
\end{figure}
\begin{figure}[!t]
\begin{center}
\includegraphics[width=8.0cm]{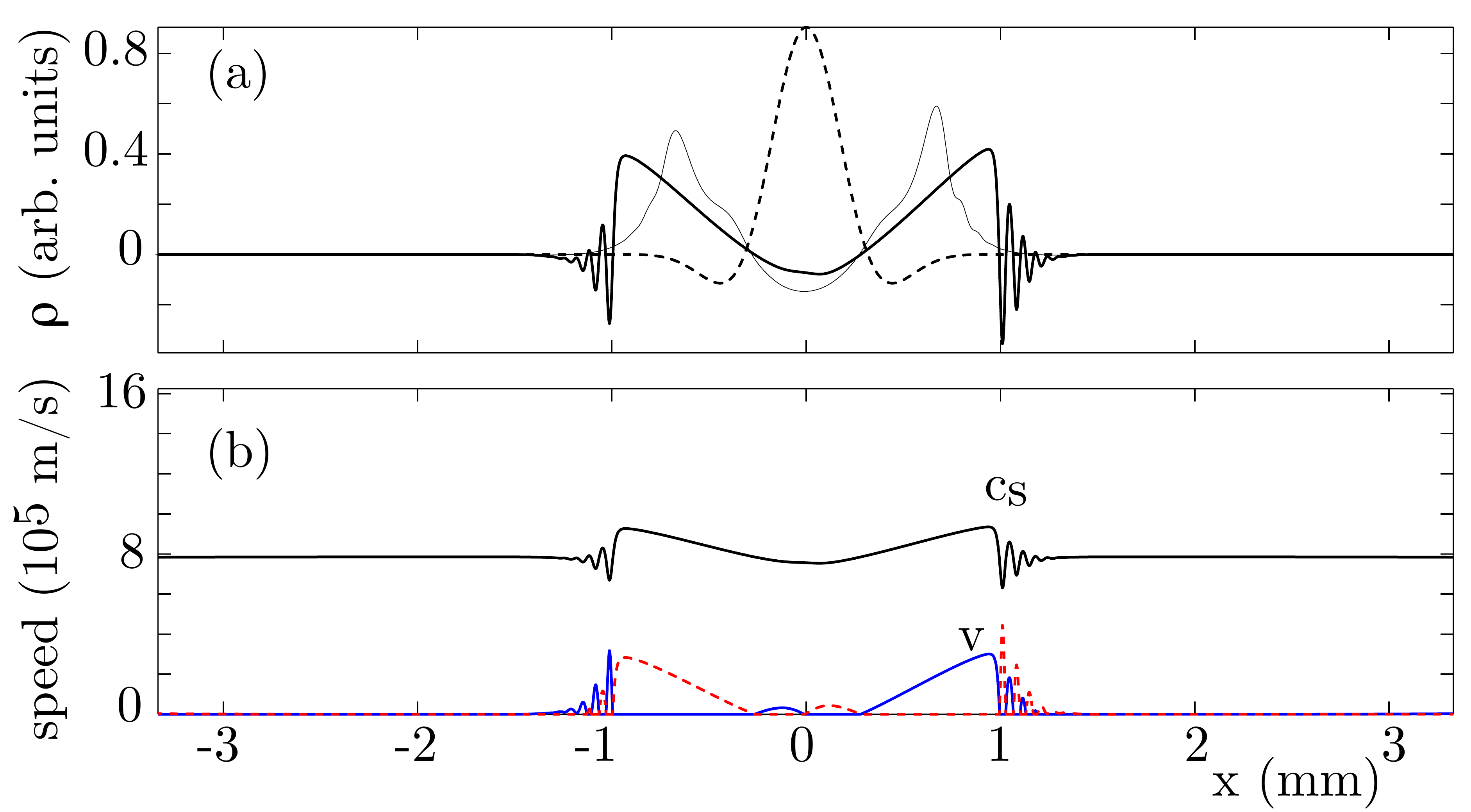}
\caption{Numerical results: (a) lineouts taken from Fig.~\ref{fig1}(b). Dashed line - input profile of the density wave. Thick solid line - density wave at z=18 cm. Thin solid line - density wave at z=18 cm for a nonlocal medium with nonlocal length $\sigma=110$ $\mu$m. (b) Sound velocity, $c_s$, in the photon fluid (black line) and fluid flow speed (blue line - positive direction speed, dashed red line - negative direction speed). \label{fig2}}
\end{center}
\end{figure}

\begin{figure}[!t]
\begin{center}
\includegraphics[width=7cm]{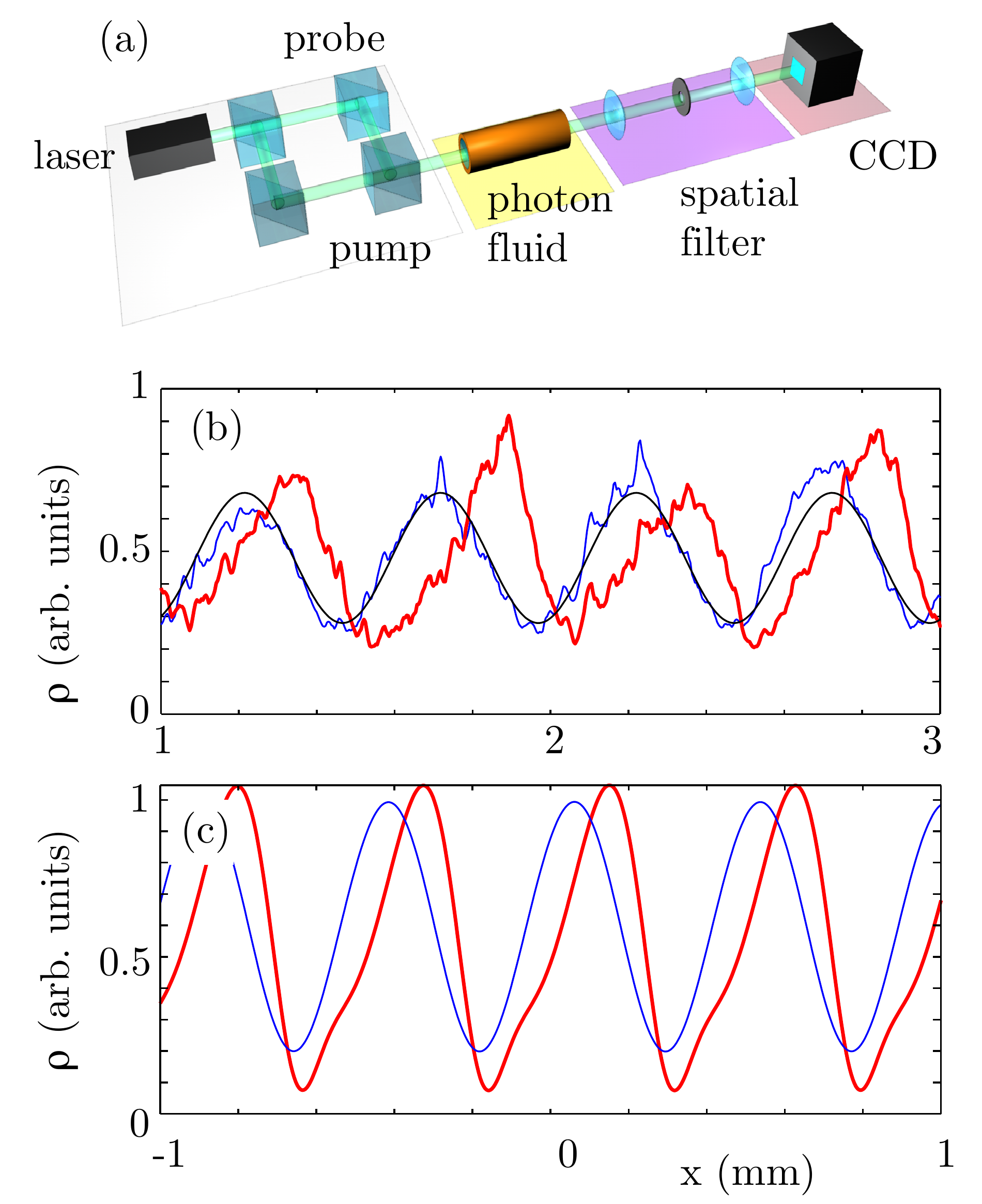}
\caption{Experimental results. (a) experimental layout. (b) measured density wave profile after spatial filtering for low powers (P=0.02 W, linear density wave propagation) and high power (P=1 W, nonlinear propagation). (c) Numerical simulation under the same conditions of the experiment, including also the nonlocal medium response, with nonlocal length $\sigma=110$ $\mu$m. 
\label{fig3}}
\end{center}
\end{figure}

{\emph{Numerical simulations:}} We numerically solved the NSE Eq.~\eqref{eq1} with input conditions similar to the experiments described below and those shown in Ref.~\cite{vocke}. A pump pulse of 532 nm and beam diameter 1 cm is propagated in a nonlinear defocusing medium with nonlinearity $n_2=10^{-8}$ cm$^2$/W. A probe beam is overlapped at the input with a small 0.2 deg angle. This forms an interference pattern, which in the photon fluid context is a density wave propagating in the background fluid (the pump beam) that has initial zero flow velocity. For a small input ampitude (less than $\sim 10$\% of the pump), the density wave splits in two counterpropagating parts (or Bogoliubov particles, see e.g. Ref.~\cite{carusotto}) as can be seen in Fig.~\ref{fig1}(a). However, for higher input powers the density waves steepen on the front edges and then breaks into higher frequency ripples, Fig.~\ref{fig1}(b). Figures~\ref{fig2}(a) and (b) show lineouts of the density wave profiles at 
 the input and output in the nonlinear acoustic propagation case together with the sound speed $c_s$ and background fluid propagation speed $v$. As can be seen, both speeds are significantly modified due to high amplitude of the propagting waves. We recall that in linear acoustic analogs, the spacetime curvature is determined by the fluid velocity and the sound speed profiles. Here the background flow is induced by the wave itself.
In other words, the initially flat spacetime geometry is curved by the waves, which are then in turn distorted by the spacetime metric.\\
{\emph{Experiments:}} We use a similar setup to Ref.~\cite{vocke} (see  Fig.~\ref{fig3}(a)). A 532 nm, CW laser beam is divided in two parts and then recombined. The more intense pump beam (1 W) forms the background fluid and the weaker probe beam propagates at a small angle so as to create, by interference, a density wave in the photon fluid. Here the probe is incident at a 0.2 deg. angle, giving rise to a density wave with a wavelength of 1 mm. This propagates in a nonlinear sample, a 21 cm long cell filled with methanol and a low concentration of graphene nanoflakes that slightly increase absorption of the pump beam. The absorbed energy is released in the form of heat, which in turn provides the defocusing nonlinearity  \cite{vocke}. After the cell, the input probe beam profile is isolated from the pump and ``idler'' beam with a spatial filter, consisting of a pinhole in the far-field of the first lens of an imaging telescope and is then imaged onto a CCD camera. The profiles of the photon fluid density wave are shown in Fig.~\ref{fig3}(b): for low amplitude input the wave propagates as a weak perturbation of the background and preserves it's initial sinusoidal shape (blue curve) but for high input amplitudes (50\% of the pump) the wave distorts and self-steepens (red curve). We note that the self-steepening observed in the experiments is somewhat less severe than that observed in the simulations. This may be ascribed to the fact that the thermal nonlinearity is intrinsically nonlocal. This has the effect of smoothening out any sharp features in the nonlinear response. For our experiment the nonlocal length $\sigma$ over which the nonlinearity extends has been measured to be $110$ $\mu$m \cite{vocke}. Numerical simulations including such a nonlocal response indeed confirm this interpretation [thin black line in Fig.~\ref{fig2}(a)].\\
{\emph{Geometrical description:}} The connection between the nonlinear fluid equations and a "dynamic" acoustic metric can be shown by writing Eqs. (\ref{eq2})-(\ref{eq4}) in terms of
the Riemann invariants, $\psi_{\pm} = v \pm 2 c_s$, where $c_s$ is the local speed of sound usually defined as $c_s^2 \equiv \frac{d P}{d \rho} = \frac{c^2 n_2}{n_0^3} \rho$. 
In terms of these variables we obtain two advection equations
\begin{equation}  
\partial_t \psi_{\pm} + c_{\pm} \partial_x (\psi_{\pm}) = 0 \, \\
\label{eq5}
\end{equation}
where $c_{\pm} = 3/4 \psi_{\pm} + 1/4 \psi_{\mp} = v \pm c_s$. Equations (\ref{eq5}) simply state that the two variables $\psi_{\pm}$ are conserved along the characteristics curves defined by $dx/dt \equiv c_{\pm}$. Therefore, in a nonlinear wave the constant quantities $\psi_{\pm}$ are propagated in spacetime at the characteristic speeds $c_{\pm} = v \pm c_s$, so they travel at the speed of sound relative to the flowing fluid. As in linear acoustic analogues, the characteristic curves demarcate the region of causally connected events. The acoustic line element can be written as
\begin{equation}
ds^2 = g_{\mu \nu} dx^{\mu} dx^{\nu} = (dx - (v + c_s) dt)(dx - (v - c_s) dt) 
\label{eq6}
\end{equation}
and thus the characteristics of the fluid equations coincide with the null geodesic defined by the acoustic metric $g_{\mu \nu}$. %Notice though that, since in the nonlinear regime $c_{\pm}= c_{\pm}(\psi_+,\psi_-)$, the metric coefficients depend on the solution.
In general, the integration of the characteristic equations is not
trivial, as Eqs. (\ref{eq5}) are weakly coupled through the coefficients $c_{\pm}$, which depend on both Riemann invariants. However the self-steepening dynamics up to the shock formation are well described by particular solutions, known as simple waves for which one of the two Riemann invariants is constant on the (x,t) plane. 
Consider an arbitrary density profile propagating on an uniform density state, $\rho_0$, with zero flow ($v=0$). After a short transient, the initial profile separates into two parts propagating in opposite directions. When the newly formed right-going and left-going profiles are well separated, they can be approximated by two independent simple waves, corresponding to $\psi_{-}$ = const for the right-going wave and to $\psi_{+}$ = const for the left-going wave. In this case, the density and the velocity are related by a specific functional relation $v(\rho) = \pm 2 (c_s(\rho) - c_s(\rho_0))$. Therefore Eqs. (\ref{eq5}), can be expressed as two independent nonlinear advection equations in just one of the hydrodynamic variables. In terms of the density we have $\partial_t \rho + (v(\rho) \pm c_s(\rho)) \partial_x \rho = 0 $
%\begin{equation}  
%\partial_t \rho + (v(\rho) \pm c_s(\rho)) \partial_x \rho = 0 \, \\,
%\label{eq7}
%\end{equation}
which can be integrated to give $x = t(v(\rho) \pm c_s(\rho)) + x_0(\rho)$,
%\begin{equation}
%x = t(v(\rho) \pm c_s(\rho)) + x_0(\rho)
%\label{eq8}
%\end{equation}
where the arbitrary function $x_0(\rho)$ depends on the intial conditions. These are the travelling wave solutions found by Riemann \cite{riemann} requiring that in Eqs. (\ref{eq2})-(\ref{eq4}) the velocity and density can be expressed as functions of one another \cite{landau}.
Notice that any point in the wave profile (i.e. a point at a given density) propagates with constant velocity $c_{\pm}(\rho) = v \pm c_s$, so the characteristics are now straight lines. Such waves can be conveniently regarded as a superposition of a density fluctuation propagating relative to the fluid, at the speed of sound and the movement of the fluid (induced by the wave itself) with velocity $v$. 
Since $d c_{+}/d \rho >0$ for right-going waves (and viceversa for left-going waves), the propagation velocity of a given point in the wave profile increases with the density: points of higher density propagate faster than points of lower density leading to self-steepening of the wave front. In a finite time the wave front will become vertical, implying an unphysical multivalued solution. The time and position in which the profile will exhibit such discontinuity is determined by the condition that $\rho(x,t)$ has an inflection point with a vertical tangent line, i.e. 
\begin{equation}
(\partial x /\partial \rho)_{t} = 0  \hspace{.8cm}
 (\partial^2 x /\partial \rho^2)_{t} = 0. 
\label{eq9}
\end{equation}
The latter conditions have a geometrical interpretation in terms of the characterisitics/null geodesics. Since the velocities $v \pm c_s$ are functions of the density, two characteristics of the same family (right-going or left-going) emanating from two spacetime points are not parallel and intersect at a finite time. As the gradient of the solution profile becomes increasingly steep, two characteristics become closer and closer. At the point of intersection, the solution is multivalued and the gradients are infinite (gradient catastophe) \cite{landau}. \\
The resulting spacetime is not null complete. In fact, the family of half null geodesics, (i.e. geodesic curves which have one endpoint and have been extended as far as possible from that endpoint), having a finite proper length, can be parametrized by a finite affine parameter \cite{wald}. Consequently, propagating waves reach a singularity in a finite time. 
The appearence of a curvature singularity can be explicitly shown considering that, for the situation above, the acoustic metric (\ref{eq6}) takes the diagonal form $ds^2 = - (3 c_s(\rho) - 2 c_s(\rho_0))^2 dt^2 + dx^2$.
In this case, it is simple to calculate the Ricci scalar, $R$, which in two dimensions fully determines the spacetime curvature. Using the fact that $c_s \propto \rho$ we obtain
\begin{equation}
R = -2 \frac{\partial_{xx}^2 \rho^{1/2}}{\rho^{1/2}} 
\label{eq11}
\end{equation}
The fact that a multi-evaluated density (or flow velocity) would imply the divergence of $R$ can be deduced from the definition of the acoustic metric, as already introduced in linear acoustic analogs. However, in linear models a singular spacetime can only be externally imposed, e.g by choosing a background with a topological defect. Here instead, a curvature singularity emerges spontaneously in a finite time, starting from a constant, homogeneous background. This is due to the \emph{nonlinear} interplay between the propagating wave and the underlying metric, whose curvature is generated by the wave itself.
Moreover, it is worth noting that $R$ has the same spatial dependence of the "quantum pressure" term in Eq. (\ref{eq3}). This term is usually neglected in the derivation of the acoustic metric in quantum fluids and, when is taken into account, is responsible for the high-energy breaking of Lorentz invariance.
Eq. (\ref{11}) shows that in our specific case, even this quantity has a geometrical interpretation, as it can be explicitly related to $R$.
On this basis, the nonlinear steepening process can be fully described in geometrical terms. A simple wave can be seen as propagating on a curved spacetime generated by the wave itself. As time evolves, the wave amplitude (i.e. the density profile) changes thus in turn modifying the curvature of the corresponding metric. In the absence of the stabilizing effect of quantum pressure, this non-linear deformation will result in a curvature singularity of the acoustic metric.\\
\begin{figure}[h!]
\begin{center}
\includegraphics[width=7cm]{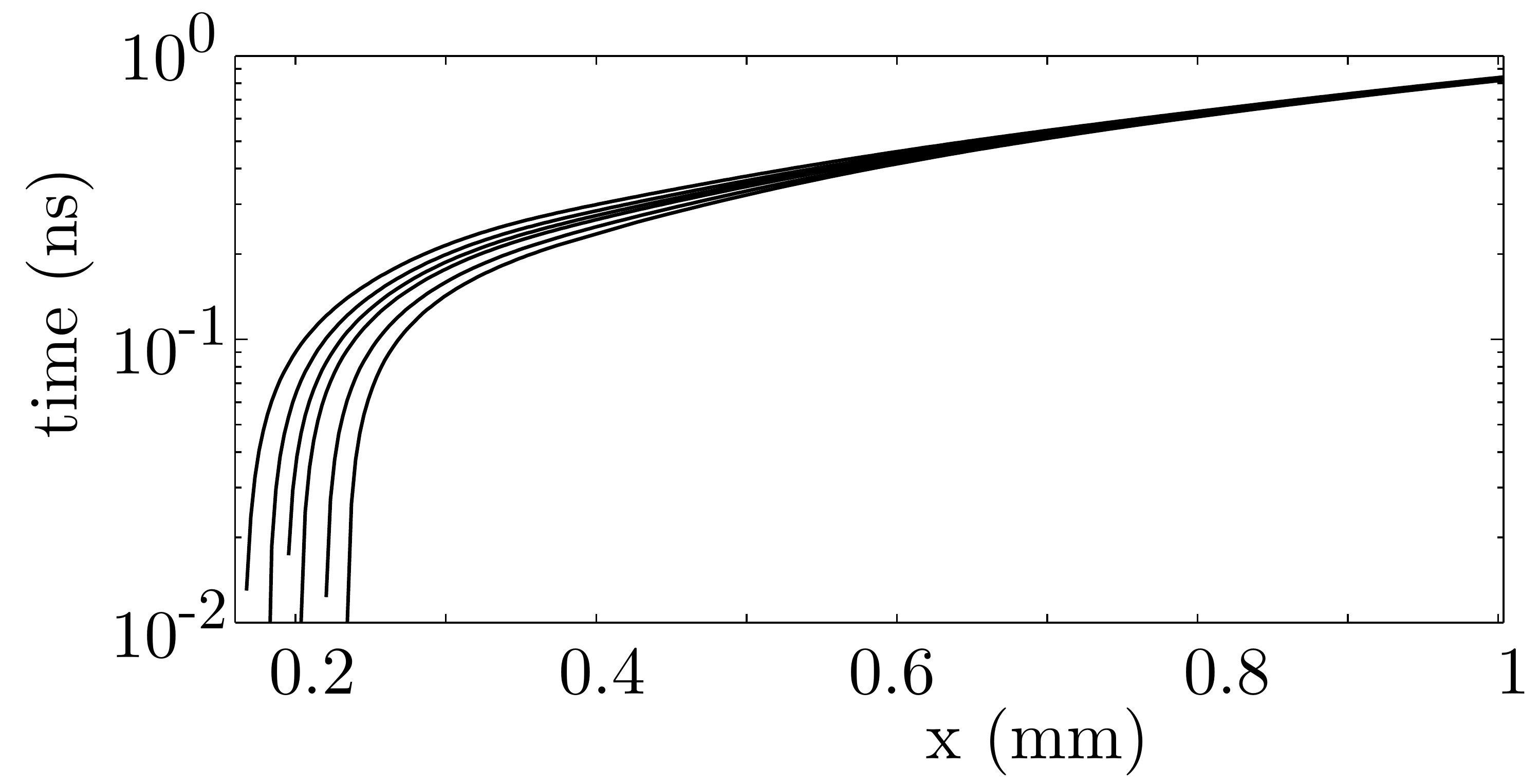}
\caption{Numerical results. Trajectories of points with the same density  generated by the right-moving high amplitude density wave shown in Fig.~\ref{fig2}(a).  \label{fig4}}
\end{center}
\end{figure}
In Fig.~\ref{fig4} we plot the numerically calculated characteristics, corresponding to the $(x,t = n_0z/c)$ trajectories of points with the same density $\rho$, for the case of the nonlinearly propagating acoustic wave of Fig.~\ref{fig2}(a). Although the presence of quantum pressure and nonlocal effects prevent the formation of a singularity, the convergence of the characteristics is a clear indication of an increasing spacetime curvature, in agreement with the ideal picture based on the emergent metric.\\
{\emph{Conclusions:}} Quantum fluids such as BECs, polariton fluids and photon fluids have been proposed as platforms to study analogue gravity effects. To date these have focused on the propagation of weak amplitude density waves on top of a given background configuration. This has led to a series of important kinematic studies, including e.g. glimpses of Hawking radiation \cite{jeff}. It is possible to extend these experimental models into the nonlinear regime where the background curved geometry determining the propagation of the waves is generated by the waves themselves. This self-interaction can thus be interpreted as the gravitational influence on the wave by its own effective metric. 
Such analogue models are truer in spirit to general relativity, where mass distributions evolve in a spacetime metric that is modified by mass itself. As mentioned before, in the presence of particular symmetries there is a precise correspondence between the gravitational field equations and the fluid dynamics. Therefore, suitably constrained photon-flows could be exploited to mimic the dynamics of gravitation black holes, spacetime singularities included \cite{cadoni}, and cosmological solutions \cite{lid,GRsolitons}.\\
 D.F. acknowledges financial support from the European Research Council under the European Unions Seventh Framework Programme (FP/2007–2013)/ERC GA 306559 and EPSRC (UK, Grant EP/J00443X/1).


\begin{thebibliography}{99}
\bibitem{br} The problem of how fluctuations influence the background has been considered in Ref. \cite{balbinot}, by expanding of the fluid equations up to second order.
\end{thebibliography}
\end{document}